\documentclass[5p,times,twocolumn]{elsarticle}
\usepackage{hyperref,amsmath}
\usepackage{xcolor}
\usepackage{bm}
\usepackage{mathtools}
\usepackage{xparse} 
\usepackage{caption}
\usepackage{subcaption}
\usepackage{booktabs, multirow}
\usepackage[colorinlistoftodos,prependcaption,textsize=tiny]{todonotes}
\usepackage{scalerel}
\usepackage{tikz}
\usetikzlibrary{svg.path}

\usepackage{hyperref}

\begin{document}

\title{Solar activity classification based on Mg II spectra:  \\ towards classification on compressed data}
\author[abto]{Sergey Ivanov}
\author[abto]{Maksym Tsizh}
\author[geneva]{Denis Ullmann}
\author[geneva,FHNW]{Brandon Panos}
\author[geneva,corres]{Slava Voloshynovskiy}
\address[abto]{Abto Software LLC, 77/T77 Heroiv UPA St., Lviv, Ukraine}
\address[geneva]{University of Geneva, Computer Science department, SIP group, Route de Drize, 7, 1227 Carouge, Switzerland}
\address[FHNW]{University of Applied Sciences and Arts Northwestern Switzerland, Bahnhofstrasse 6, 5210 Windisch, Switzerland}
\address[corres]{Corresponding author: svolos@unige.ch; Tel.: +41(22)-379-01-58}

\begin{abstract}
Although large volumes of solar data are available for investigation and study, the vast majority of these data remain unlabeled and are therefore not amenable to modern supervised machine learning methods. Having a way to accurately and automatically classify spectra into categories related to the degree of solar activity is highly desirable and will assist and speed up future research efforts in solar physics. At the same time, the large volume of raw observational data is a serious bottleneck for machine learning, requiring powerful computational means that are not at the disposal of many laboratories. Additionally, the raw data communication imposes some restrictions on real time data observations and requires considerable bandwidth and energy for the onboard solar observation systems. To cope with the above mentioned issues, we propose a framework to classify solar activity on compressed data. To this end, we used a labeling scheme from a pre-existing vector quantization technique in conjunction with several machine learning algorithms to categorize spectra of singly-ionized magnesium Mg II measured by NASA's Interface Region Imaging Spectrograph small explorer satellite IRIS into several groups characterizing solar activity. Our training dataset is a human annotated list of 85 IRIS observations containing 29097 frames in total or equivalently 9 million Mg II spectra. The annotated types of Solar activities are: active region, pre-flare activity, Solar flare, Sunspot and quiet Sun. We used the vector quantization to compress these data and to reduce its complexity before training classifiers. From a host of classifiers, we found that the XGBoost classifier produced the most accurate results on the compressed data, yielding over a 95\% prediction rate, and outperforming other ML methods like convolution neural networks, K-nearest neighbors, naive Bayes classifiers and support vector machines. A principle finding of this research is that the classification performance on compressed and uncompressed data is comparable under our particular architecture, implying the possibility of large compression rates for relatively low degrees of information loss.
\end{abstract}

\maketitle

\section{Introduction}
The prediction and classification of Solar activity and Solar phenomena in general with methods of machine learning (ML) has become an active field of research in astrophysics, having already dozens of papers devoted to this subject, and widely cited in \cite{McCloskey_2016} and \cite{Barnes_2016}. 
With an exponential growth of solar data, fast algorithms that can guide researchers to specific solar activity becomes a very useful tool enabling exploration of new phenomena.

Some previous research efforts in this direction include feature recognition by \cite{Turmon10} using a maximum aposteriori (MAP) technique on line-of-sight magnetic flux images, a highly successful application of deep convolutional neural networks by \cite{Armstrong19} to classify a range of different solar features including amongst others, filaments, prominences and flare ribbons, as well as a solar flare classification and prediction paper by \cite{Jiao19}.

The majority of effort within the literature has been devoted to the prediction of solar flares, rather than to the classification of different types of solar activity. These efforts are most often based on photospheric vector-magnetic field data from the Solar Dynamics Observatory's on-board Helioseismic and Magnetic Imager (HMI) \cite{HMI_2012}. A comprehensive overview of flare prediction papers based on ML models can be found in \cite{Chen19}, where the authors themselves use HMI magnetic data coupled with soft X-ray data from the Geostationary Operational Environmental Satellites (GOES). Amongst the most recent publications: \cite{Campi19} performed a feature ranking of 171 flare-prediction features with a hybrid LASSO and random forest classifiers and \cite{Liu19} used a long short-term memory network (LSTM), to predict the occurrence of different flare classes within a 24 hours time frame. 

As reported in \cite{Florios_2018} and \cite{Kontogiannis_2018}, magnetic fields are natural predictors for solar flares, since magnetic reconnection is the source of eruptive events on the Sun. Nevertheless, instead of defining solar activity from a magnetic perspective, we have chosen to categorize the activity by monitoring directly observed spectral responses. Such input data would be model-free, and not rely on tenuous extrapolations of photospheric magnetic data into the corona, like the works on spectropolarimetric diagnostics and inversion methods using ML proposed by \cite{Asensio_Ramos_2012, Mili__2020, Osborne_2019}. Moreover, there exists a previously developed algorithm for data compression and clusterization as suggested in \cite{Panos18} that will make this classification study considerably simpler. This algorithm is given in the \emph{irisreader} python library (\emph{assign\_mg2k\_centroids} function) and assigns each observed spectra to their nearest neighbor centroid found by \cite{Panos18}.

\begin{figure}
\begin{center}
\includegraphics[trim = 0mm 71mm 98mm 0mm, clip, width=\linewidth]{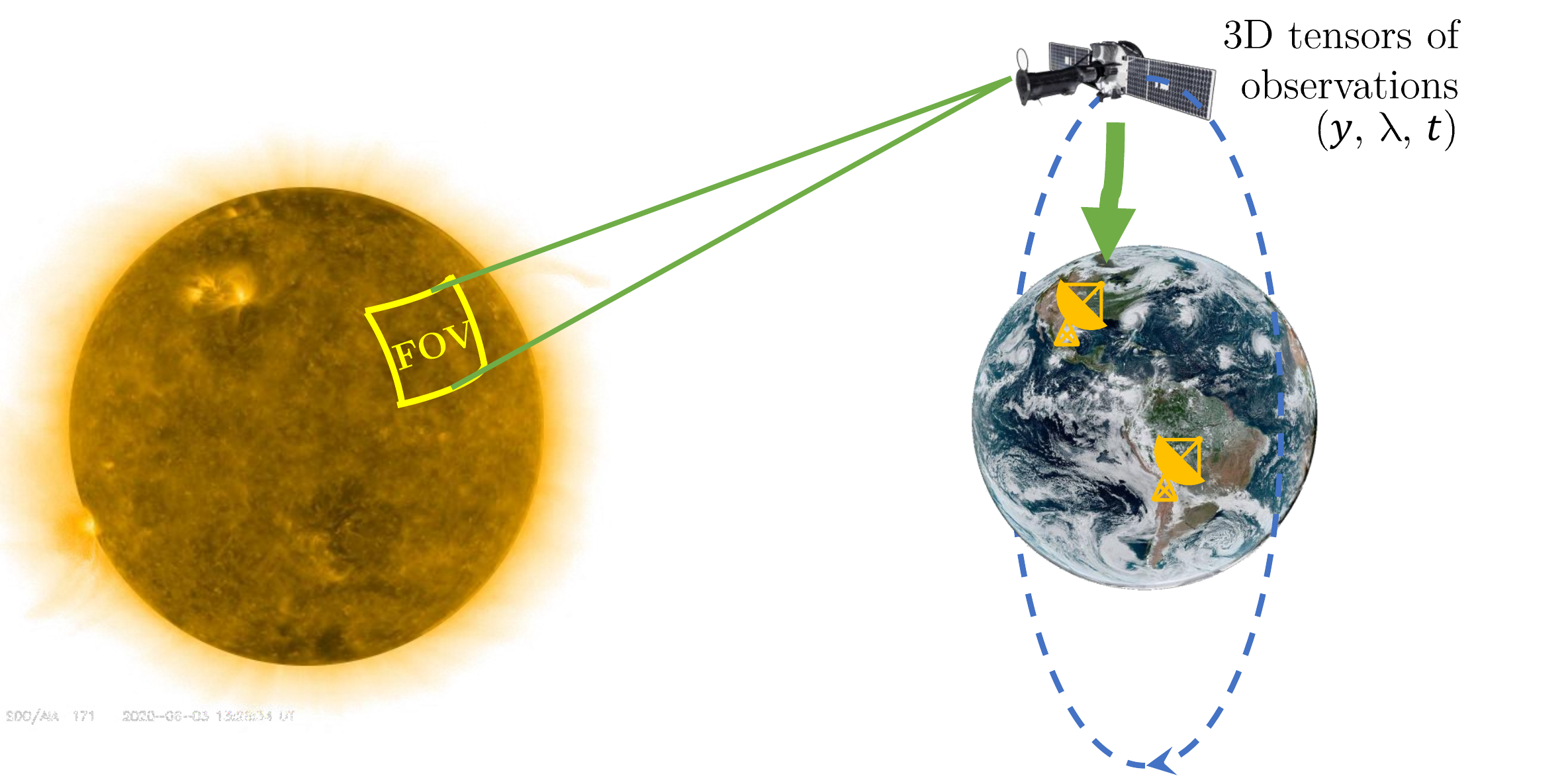}
\end{center}
\caption{Cartoon demonstrating the information Bottlenecks of IRIS, including some of the challenges of data acquisition and transmission: 1) Even if the field of view (FOV) only partially covers the solar surface. 2) It is impossible to store all the high definition 3D tensors, because of limited onboard memory and transmission to the Earth can only be performed at certain times and positions along the satellites telemetry. Sun and Earth image credits: NASA's Goddard Institute for space studies.}
\label{fig:Compression-formulation-challenge}
\end{figure}

\begin{figure}
\begin{center}
\includegraphics[trim = 0mm 88mm 107mm 0mm, clip, width=\linewidth]{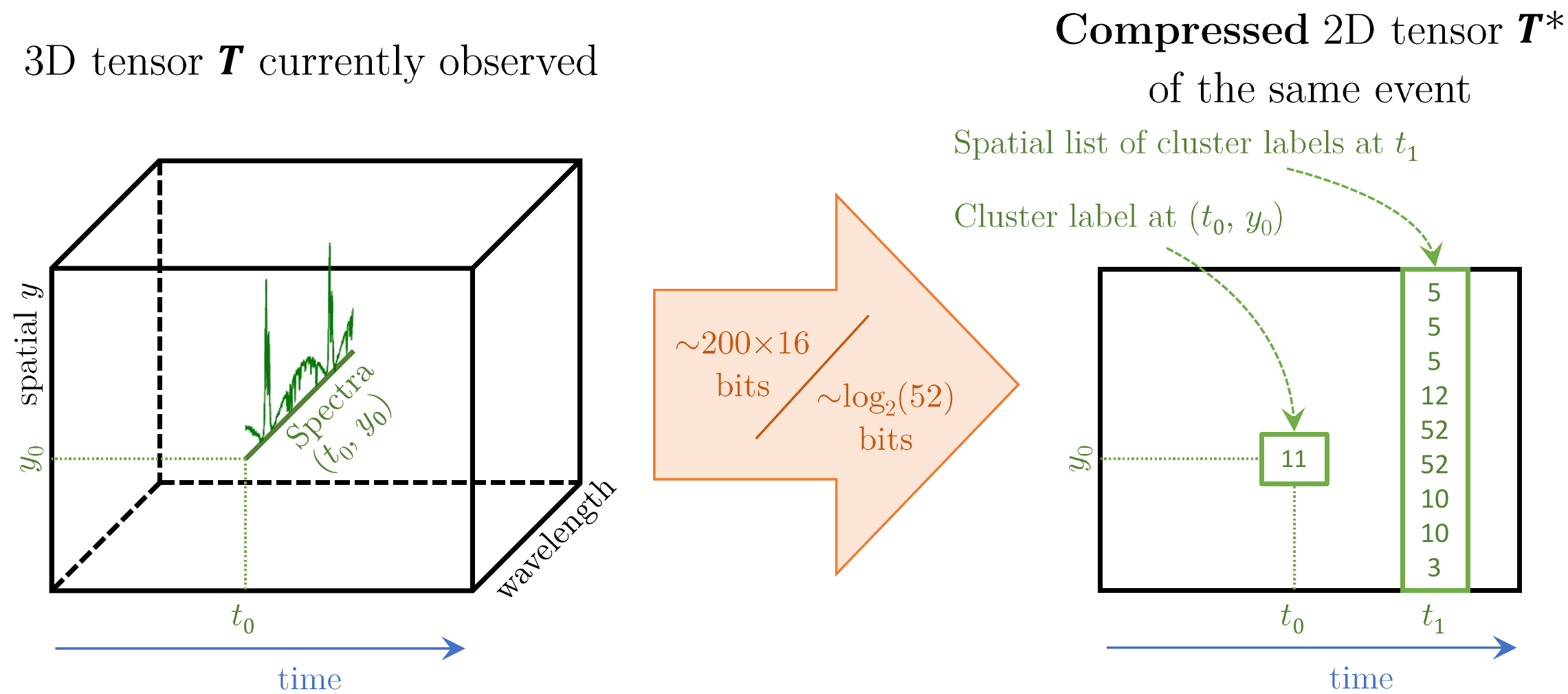}
\end{center}
\caption{Diagram of proposed data compression from 3D tensor to 2D matrix representation. Spectral lines of length 240 and 16 bits representation per wavelength are encoded into 52 integers representing 52 cluster centers. The compression ratio is $\frac{240\times16}{\log_2(52)} = 674$:1.}
\label{fig:Compression-rate}
\end{figure}

The spectral data for this study is from the NASA's Interface  Region  Imaging Spectrograph IRIS satellite \cite{DePontieu2014}, which faces several challenges of data acquisition and transmission, some of which are stated and described in Figure \ref{fig:Compression-formulation-challenge}. The IRIS imager records images and spectral data in raw format with a 16 bits representation per sample. One hour of spectral data alone can come to approximately 350 Megabytes. The IRIS imager usually performs the observation of a pre-selected $175\times175~\text{arcsec}^2$ solar region over a time period of a few hours. The accumulated data are then communicated to a ground base station to free the pre-processing and storage facilities for the next observation campaign. This creates a bottleneck in spatial and temporal observations. Other satellite based missions face similar data challenges, therefore, the model presented in this work will be of interest for the planning of future missions with the increased coverage of field of view, real time communications and more flexible data communication to the Earth base stations.
As a consequence, unlike the previously reported raw data classification methods, we consider as an option for future missions to have a low-complexity data compression on board as shown in Figure \ref{fig:Compression-rate}, and sequentially we analyze a possibility of classification based on \emph{compressed} data.\\

This work has the following structure. In section \ref{IRIS_data}, we describe our data set and pre-processing pipeline used to make the spectral data compatible for machine learning purposes. In section \ref{Proposed_approuch_section}, we introduce the mathematical formulations of our approach, while in sections \ref{Spectral_compression_section} and \ref{Classifier_section} we lay out the particulars of the compression and classifier schemes outlined theoretically in the previous section. In section \ref{Results_section}, we present our results, compare the performance of the classifiers on compressed and raw spectral data, and present a single flare case study, while addressing the limitations  and possible extensions of our work. The paper is finalized with a conclusion in section \ref{Conclusion_section}.

\section{IRIS data}
\label{IRIS_data}
\subsection{IRIS observations}
The data for this study has been acquired by NASA's IRIS small explorer spacecraft presented in \cite{DePontieu2014}. IRIS can record spectral data in both the near-ultraviolet (NUV) and far-ultraviolet (FUV) passbands of several important chromospheric and transition region lines. At the same time, IRIS can take contextual images of the solar atmosphere via the use of a small filed of view $175\times175~\text{arcsec}^2$ slit-jaw imagery (SJI). IRIS has several operational modes, where scientists can plan mission campaigns with cadences, rasters and exposure times to suit their specific science goals, such that each event has different observation settings. The spatial location, orientation, position of the slit on the image, cadence of observations, wavelengths, range of observations may vary between events and sometimes also during one single event. The dexterity of the instrument introduces a unique hurdle for machine learning projects, where the inherent heterogeneity of the data must first be homogenised. Since its launch in 2013, IRIS has observed hundreds of large flares, and played a crucial role in our current understanding and parameterization of flare dynamics. The Mg II resonant lines with core vacuum wavelengths at $2803.52$ and $2796.34$ $\text{\AA}~$ are two of the strongest lines in IRIS's NUV passband. They have a formation height that extends across the entire chromosphere, and in combination with a triplet of close subordinate lines, have proven to be a particularly rich source of diagnostic information, allowing us to deduce velocities, opacity's, temperatures and densities within the chromosphere \cite{Leenaarts_2013}. The use of Mg II profiles for the classification of different types of solar activity is well founded, with distinctions in profile shapes between quiet Sun and sunspot umbra documented some 40 years ago by \cite{Gurman_1984}. More recent publications have stressed the dramatic difference between quiet Sun profiles and flaring profiles \cite{Kerr}.

\subsection{Pre-prossessing}
An additional layer of data cleaning was applied to the standard level2 IRIS data product \cite{IRIS_l2}. This step was necessary to make the data compatible with our research aims and machine learning techniques. The Mg II lines were first extracted from a window between 2794.14 and 2805.72 $\text{\AA}$. Spectra with data counts below 7 DN/s\footnote{ DN stands for "Digital Number" and is the value given by the sensor. As announced in \cite{DePontieu2014} is calibrated such that there are about 18 photons captured per DN.} were then removed on account of bad signal to noise ratios. Spectra with missing data in the form of large negative values as well as overexposed spectra with more than 5 consecutive points at the same intensity were also removed. All spectra were then interpolated onto the same wavelength grid of 216 points. This is an important step since many machine learning techniques, including the ones used in this paper, require the input data to have the same dimensionality. Finally, the spectra were normalized by their maximum value so that spectral global shape, but not its intensity, becomes the main feature for our methods. Although this normalization leads to a loss of information, the centroids found in \cite{Panos18} are normalized, which in turns implies that the intensity has no influence on the compression scheme that assigns each spectrum to their nearest neighbor centroid. In the same manner as has been suggested by \cite{Panos18}, we assume here that the intensity of an event and its activity are two independent things, and that small or big solar flares seam to share the same physics.

\section{Theoretical proposition}
\label{Proposed_approuch_section}
The proposed approach is inspired by an information bottleneck (IB) principle proposed by \cite{tishby2000information} in application to supervised deep network classification. The IB terminology refers to the mutual information between two random variables $\bm{X}$ and $\bm{Y}$, and defined in Shannon's information theory \cite{ShannonIT} by:
\begin{align}
I(\bm{X}; \bm{Y}) &= \iint_{(x,y)}p_{\bm{X}, \bm{Y}}(x,y)\log_2\frac{p_{\bm{X}, \bm{Y}}(x,y)}{p_{\bm{X}}(x)p_{\bm{Y}}(y)}dxdy\\
 &= \mathbb{E}_{p_{\bm{X}, \bm{Y}}}\left[\log_2\frac{p_{\bm{X}, \bm{Y}}}{p_{\bm{X}}p_{\bm{Y}}}\right].
\end{align}
The mutual information quantifies a common information between $\bm X$ and $\bm Y$ by comparing the joint distribution $p_{\bm{X}, \bm{Y}}$ and the product of the marginal distributions $p_{\bm{X}}p_{\bm{Y}}$; when $\bm{X}$ and $\bm{Y}$ are independent, $I(\bm{X}; \bm{Y})=0$ and high value of $I(\bm{X}; \bm{Y})$ means that $\bm{X}$ and $\bm{Y}$ share a lot of common information.

\begin{figure}
\begin{center}
\includegraphics[trim = 0mm 20mm 12mm 0mm, clip, width=\linewidth]{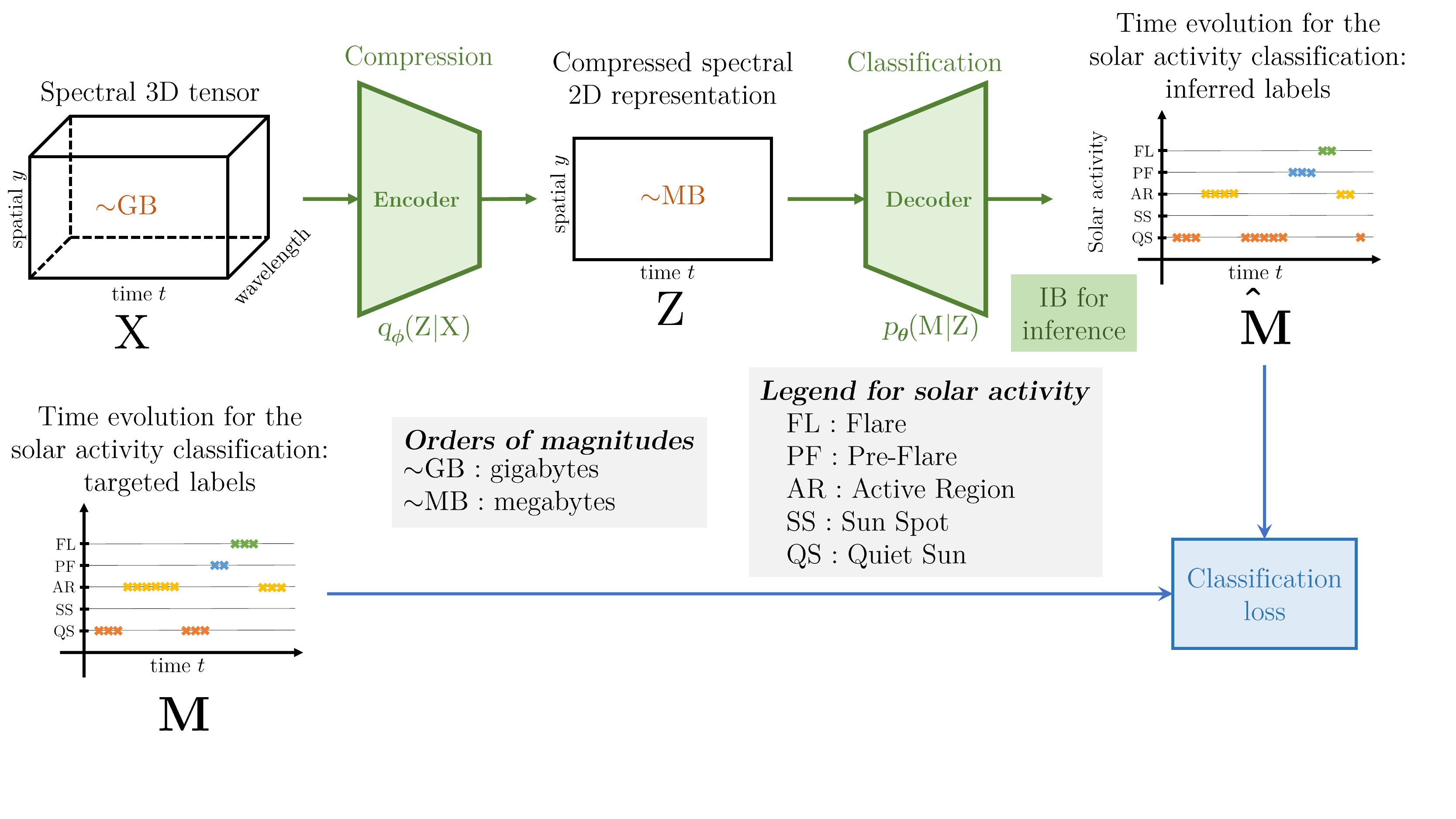}
\end{center}
\caption{IB framework in application to spectral data classification. The encoder and decoder (classifier) are jointly optimized. The solar activity is represented by a class label $\bm{M}\in\mathcal{M}=\left\{QS, SS, AR, PF, FL\right\}$.}
\label{fig:IB-fmw-spectra}
\end{figure}

The IB inference model assumes that the parametrized encoder $q_{\boldsymbol{\phi}}(\bm{Z}|\bm{X})$ compresses the input $\bm{X}$ to a latent compressed representation $\bm{Z}$ containing all sufficient statistics for the reliable \emph{classification} of labels $\bm{M}$ by the parametrized decoder/classifier $p_{\boldsymbol{\theta}}(\bm{M}|\bm{Z})$ as shown in Fig.\ref{fig:IB-fmw-spectra}.

The parameters of encoder $\boldsymbol{\phi}$ and decoder $\boldsymbol{\theta}$ are optimized jointly based on the minimization of the following cost function:
\begin{align}
\mathcal{L}(\boldsymbol{\phi}, \boldsymbol{\theta}) &= I_{\boldsymbol{\phi}}(\bm{X}; \bm{Z}) - \beta I_{\boldsymbol{\phi}, \boldsymbol{\theta}}(\bm{Z}; \bm{M}),\\
(\hat{\boldsymbol{\phi}}, \hat{\boldsymbol{\theta}}) &= argmin_{\boldsymbol{\phi}, \boldsymbol{\theta}}\mathcal{L}(\boldsymbol{\phi}, \boldsymbol{\theta}),
\label{eq:IBcmp2}
\end{align}

where $I_{\boldsymbol{\phi}}(\bm{X}; \bm{Z})$ denotes the mutual information between $\bm{X}$ and $\bm{Z}$ at the encoder and $I_{\boldsymbol{\phi}, \boldsymbol{\theta}}(\bm{Z}; \bm{M})$ denotes the mutual information between $\bm{Z}$ and $\bm{M}$ at the decoder and $\beta$ is a Lagrangian multiplier. Thus, the optimal solution is a trade-off between the compression and classification.

\begin{figure}
\begin{center}
\includegraphics[trim = 0mm 39mm 15mm 0mm, clip, width=\linewidth]{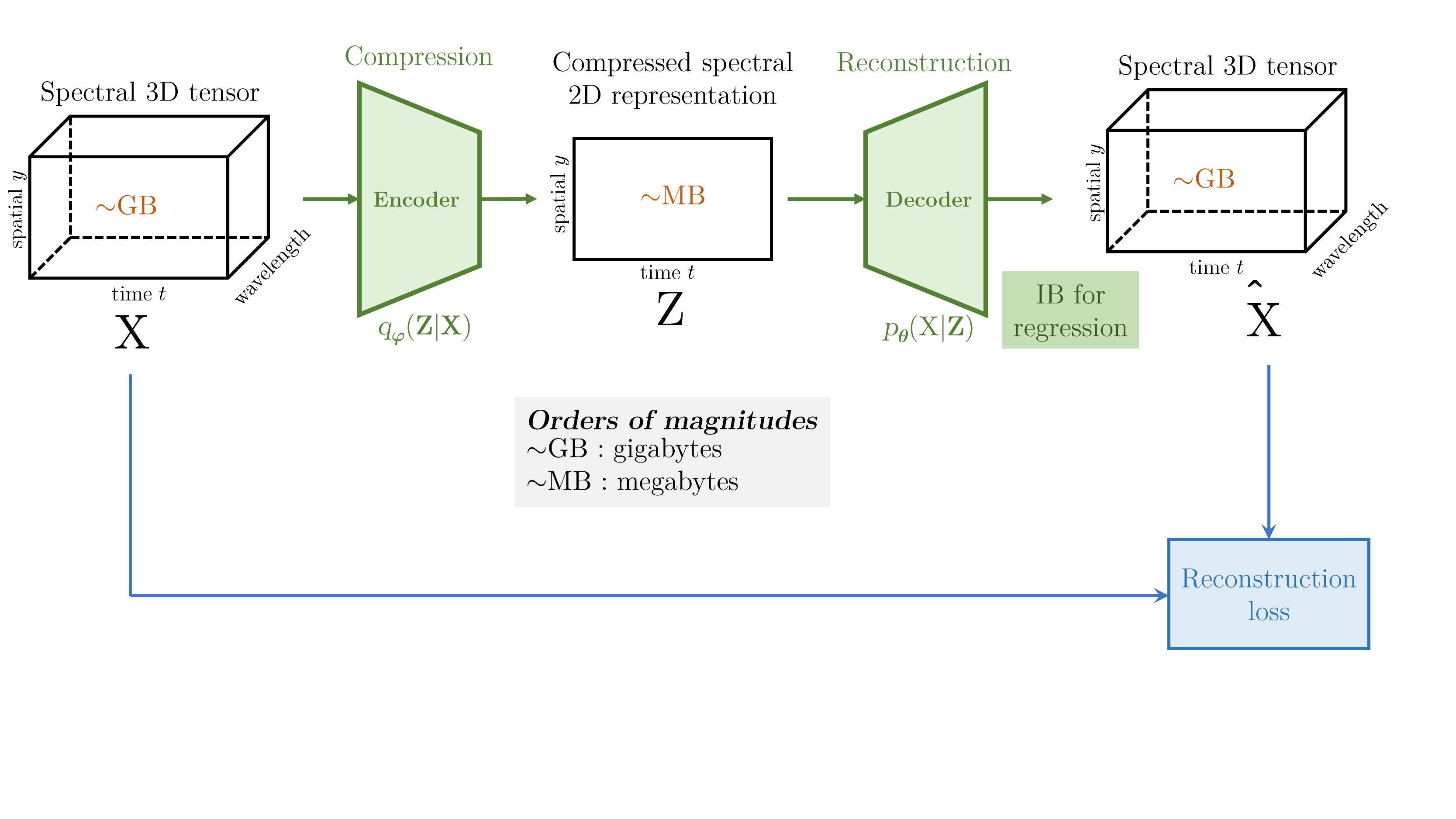}
\end{center}
\caption{IB framework in application to spectral data compression. The encoder and decoder (reconstructor) are jointly optimized.}
\label{fig:IB-fmw-spectra-comp}
\end{figure}

At the same time, the IB data compression formulation corresponds to the case, when both the encoder and decoder are jointly optimized to ensure the \emph{reconstruction} of data from the compressed latent representation as shown in Fig.\ref{fig:IB-fmw-spectra-comp}.

The IB compression formulation corresponds to the following optimization problem:
\begin{align}
\mathcal{L}(\boldsymbol{\phi}, \boldsymbol{\theta}) &= I_{\boldsymbol{\phi}}(\bm{X}; \bm{Z}) - \beta I_{\boldsymbol{\phi}, \boldsymbol{\theta}}(\bm{Z}, \bm{X}),\label{eq:IBcmp3}\\
(\hat{\boldsymbol{\phi}}, \hat{\boldsymbol{\theta}}) &= argmin_{\boldsymbol{\phi}, \boldsymbol{\theta}}\mathcal{L}(\boldsymbol{\phi}, \boldsymbol{\theta}).\label{eq:IBcmp4}
\end{align}

When the encoder and decoder are deterministic mappers and the reconstruction loss is a mean square error (MSE), as it was shown by \cite{voloshynovskiy2019information} and Eq.\ref{eq:IBcmp3} reduces to a well-known vector quantization (VQ) problem. In this case, the compressed representation is $\bm{Z}=f_{\boldsymbol{\phi}}(\bm{X})$ and $f_{\boldsymbol{\phi}}(.)$ is a deterministic encoder and $\hat{\bm{X}}=g_{\boldsymbol{\theta}}(\bm{Z})$ is a deterministic decoder and
\begin{equation}
\mathcal{L}(\boldsymbol{\phi}, \boldsymbol{\theta}) = H\left(f_{\boldsymbol{\phi}}(\bm{X})\right) + \beta\ \mathbb{E}_{p_{\mathcal{D}}(\bm{x})}\left[\left\|\bm{X}-g_{\boldsymbol{\theta}}\left(f_{\phi}(\bm{X})\right)\right\|_2^2\right],\label{eq:IBdet}
\end{equation}
where $H(.)$ denotes the entropy and $\mathbb{E}_{p_{\mathcal{D}}(\bm{x})}\left[.\right]$ stands for the expectation operator with respect to data distribution $\bm{x}\sim p_{\mathcal{D}}(\bm{x})$. The obtained solution of Eq.\ref{eq:IBdet} is a rate-distortion function \cite{Cover}. If one fixes some rate $R_{\bm{Z}}=H(f_{\boldsymbol{\phi}}(\bm{X}))$ for the latent space representation, the obtained distortion of compression corresponds to the MSE in the second term of Eq.\ref{eq:IBdet}. It should be pointed out an essential difference in optimal representation of latent space data between the problems of Eq.\ref{eq:IBcmp2} and Eq.\ref{eq:IBcmp4}, e.g., between the classification and compression cases. In the case of compression, the encoder retains an essentially larger amount of information in comparison to the classification case to ensure the desired level of reconstruction. It is explained by higher entropy of data in comparison to the entropy of labels.

\begin{figure}
\begin{center}
\includegraphics[trim = 0mm 20mm 15mm 0mm, clip, width=\linewidth]{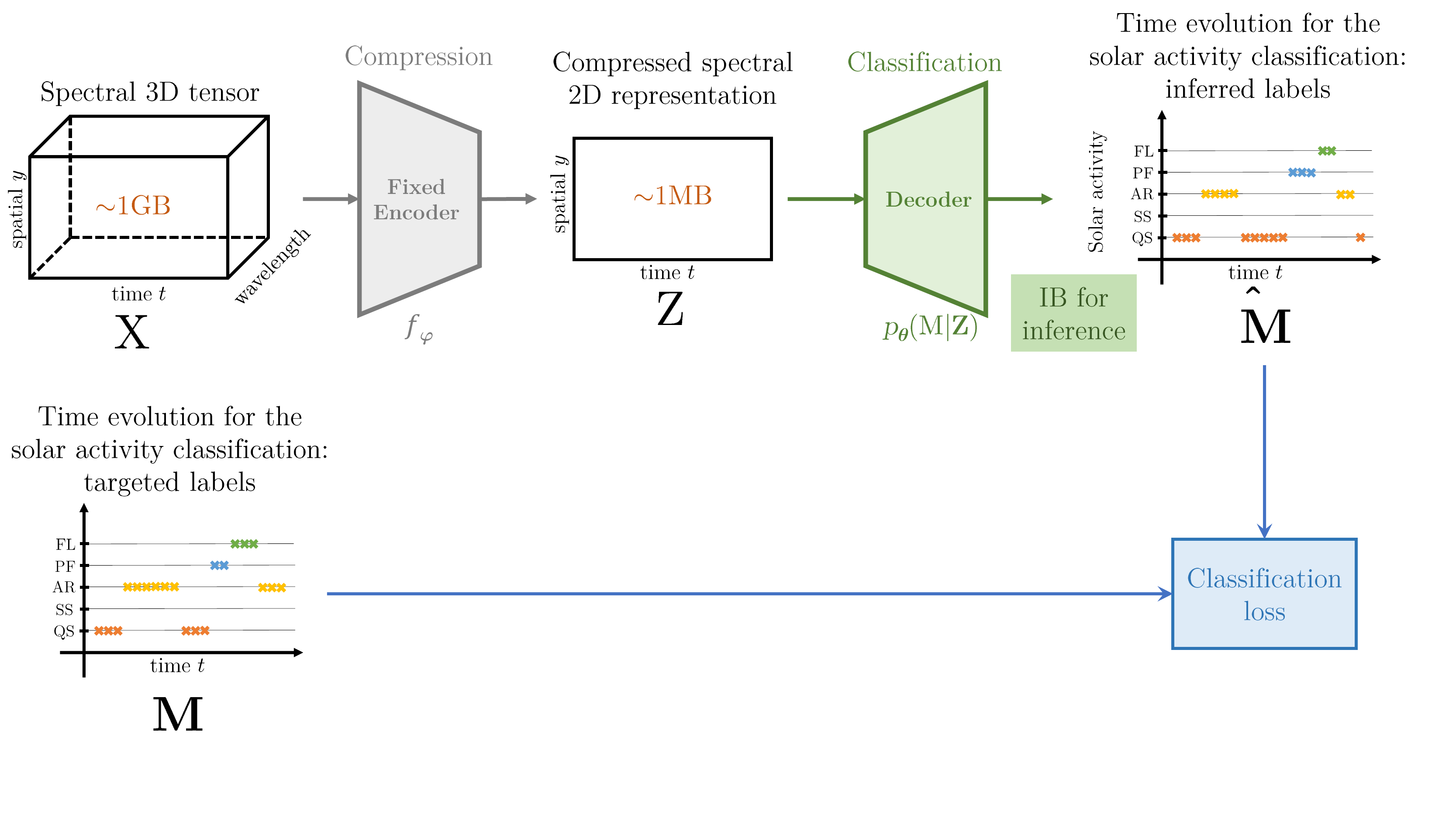}
\end{center}
\caption{Considered classification system based on compressed data. The encoder is pre-trained under the compression IB and fixed while the decoder is a trained classifier.}
\label{fig:IB-fmw-considered}
\end{figure}

In this work, we consider a practical solution based on a hybrid system. We will assume that the compression system can compress 3D data tensor to a 2D spectral representation. These 2D data will serve as an input to the decoder/classifier. A practical advantage of the considered system comes from the fact that the 3D data tensor $\bm{X}$ is compressed to the latent representation $\bm{Z}$ in the classification setup of Fig.\ref{fig:IB-fmw-spectra}. The original 3D data tensor cannot be reconstructed anymore, since $\bm{Z}$ contains only class relative information. However, many practical applications require such a compressed representation that can be further suitable for both visual analysis and classification. At the same time, if $\bm{Z}$ is a representation of compression IB formulation as in Eq.\ref{eq:IBcmp4}, the original data $\bm{X}$ can still be reconstructed from $\bm{Z}$ with a controlled fidelity. That is why we will assume that the encoder in our system is fixed and pre-trained based on the compression setup of Eqs.\ref{eq:IBcmp3} and \ref{eq:IBcmp4} or practically \ref{eq:IBdet}, as shown in Fig.\ref{fig:IB-fmw-considered} and the latent space representation will be further used for the classification.

The decoder represents a trainable classifier $p_{\boldsymbol{\theta}}(\bm{M}|\bm{Z})$ producing the labels from the compressed representation $\bm{Z}$.

Therefore, the research question is to investigate the performance of such a system under the different models of classifiers.

\section{Spectral compression}
\label{Spectral_compression_section}
\begin{figure}[htb]
\centering\includegraphics[trim={0cm 0cm 0cm 0cm},clip, width=0.5\textwidth]{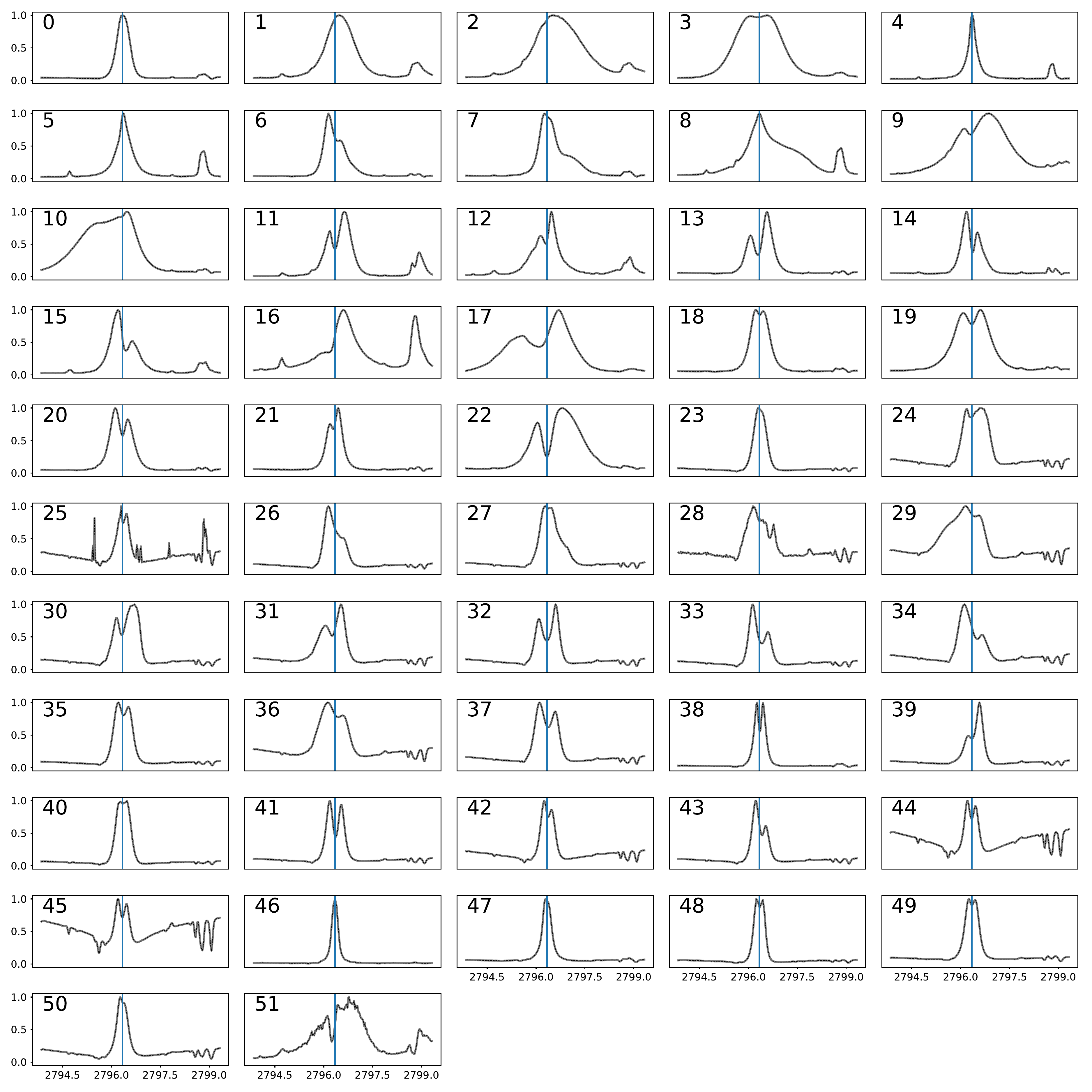}
\caption{Centroids used as a code book in the compression $X\to\tilde{X}$. The y-axis of each subplot indicates the normalized intensity, while the x-axis is wavelength in Angstroms. The blue vertical lines indicate the location of the k-core at 2796.34 Å. Most centroids capture a particular facet of the flaring process, profiles such as 44 and 46 are related to quiet Sun and sunspot emissions respectively.}
\label{Centroids}
\end{figure}

The Mg II resonant lines have a frequency dependant source function that has a complex interplay with the optical depth unity along the line of sight, resulting in a large variety of possible line shapes. Although this complexity is directly responsible for the production of excellent diagnostics, it makes the analysis of large volumes of data extremely difficult. \cite{Panos18} circumvented this problem by lowering the resolution of the data in a controlled way, while still retaining it acceptable for diagnostic. This was achieved by employing a classical clustering algorithm known as the k-means algorithm, which partitions the data into a predefined $k$ number of groups, such that the final partitioning minimizes the within cluster variance and represents the compression scheme addressed in the present work. Each partition or group can then be approximated by the group mean, often referred to as the representative profile or centroid. A table of the 52 centroids used in this study can be found in Figure \ref{Centroids}, and provides us with a dictionary that can be used with the nearest neighbors approach to transform the high dimensional sequence of spectra $\{\bm{X}_i\}_{i=1}^{N_{\bm{X}}}$ into the corresponding list of labels $\{l_i\}_{i=1}^{N_{\bm{X}}}$, i.e., $\bm{X}_i\in\mathbb{R}^{626}\to l_i\in\left\{1,\dots,52\right\}$, with $N_{\bm{X}}$ to be the temporal size of $\bm{X}$, and where each spectrum is assigned the label of the centroid which is the closest to. It must be noted that the 52 centroids seen in Figure \ref{Centroids} were designed with an emphasis on the spectral variations within flares, therefore, they may not serve as an ideal basis for the problem at hand. Nevertheless, there are components of this table endemic to all 5 solar regions under the investigation in this paper. 

\begin{figure}[htb]
\centering\includegraphics[trim={0cm 0cm 0cm 0cm},clip, width=0.45\textwidth]{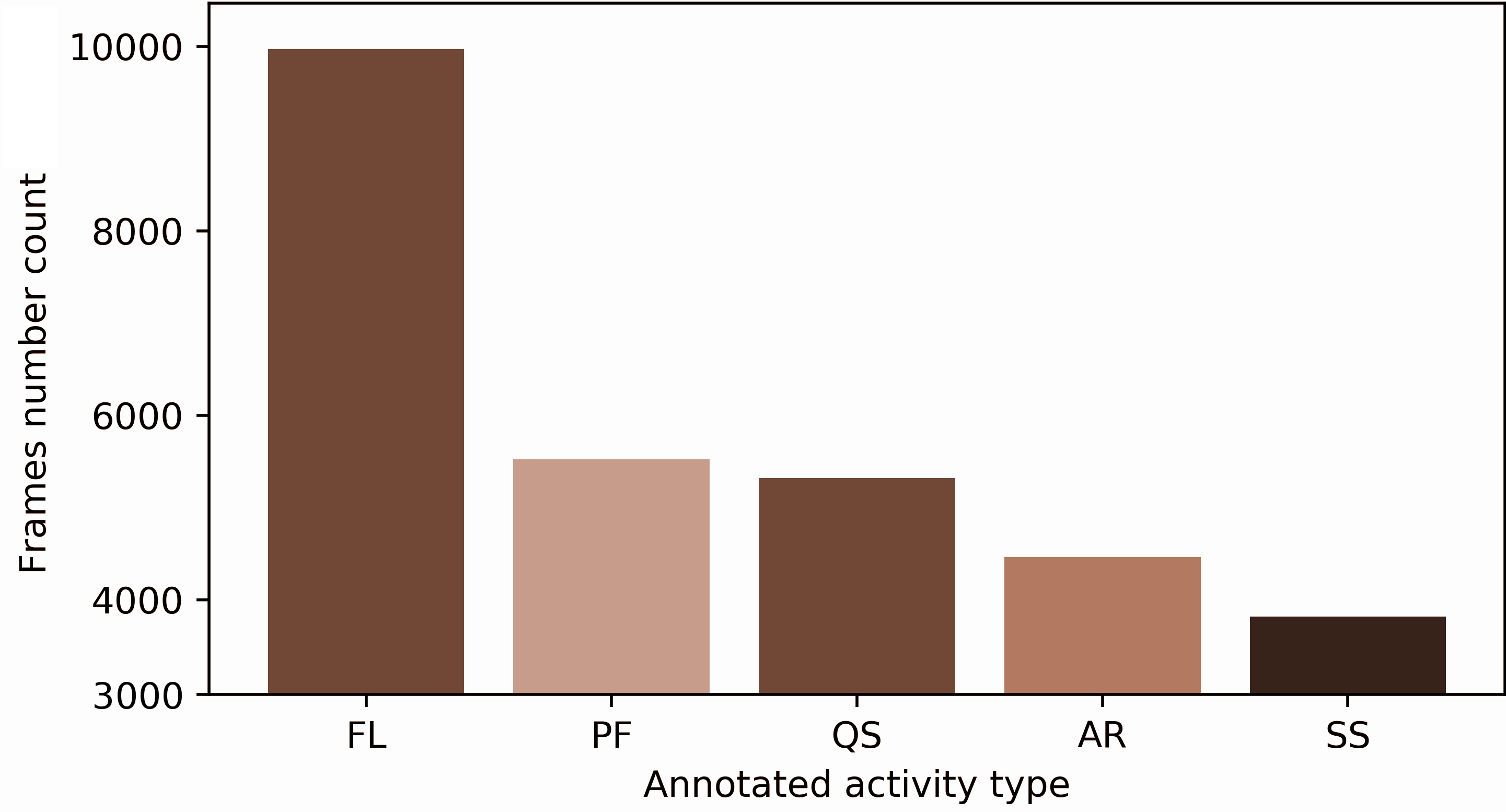}
\caption{Distribution of frames by the type of annotated activity}
\label{classes_count}
\end{figure}

To train our ML algorithm we have used the dataset of 85 IRIS observations, each of them was annotated to have only one type of activity on it. There are 29097 frames (photos) of Sun in total in these observations.  We assume there are 5 types of Solar activity. One can associate these observation with: active region (AR), pre-flare activity (PF), Solar flare (SF), Sunspot (SS), quiet Sun (QS). The distribution of frames by the activity types is given in Fig.\ref{classes_count}.
We expect that some of the centroids of Fig. \ref{Centroids} from the clustering method are endemic to each of the 5 types of solar activity. We want to use a ML classification technique that can deduce at each time the activity class from the underlying statistics of these centroids.

\begin{figure*}[thb]
\centering\includegraphics[trim={0cm 2.3cm 1cm 0cm},clip, width=\textwidth]{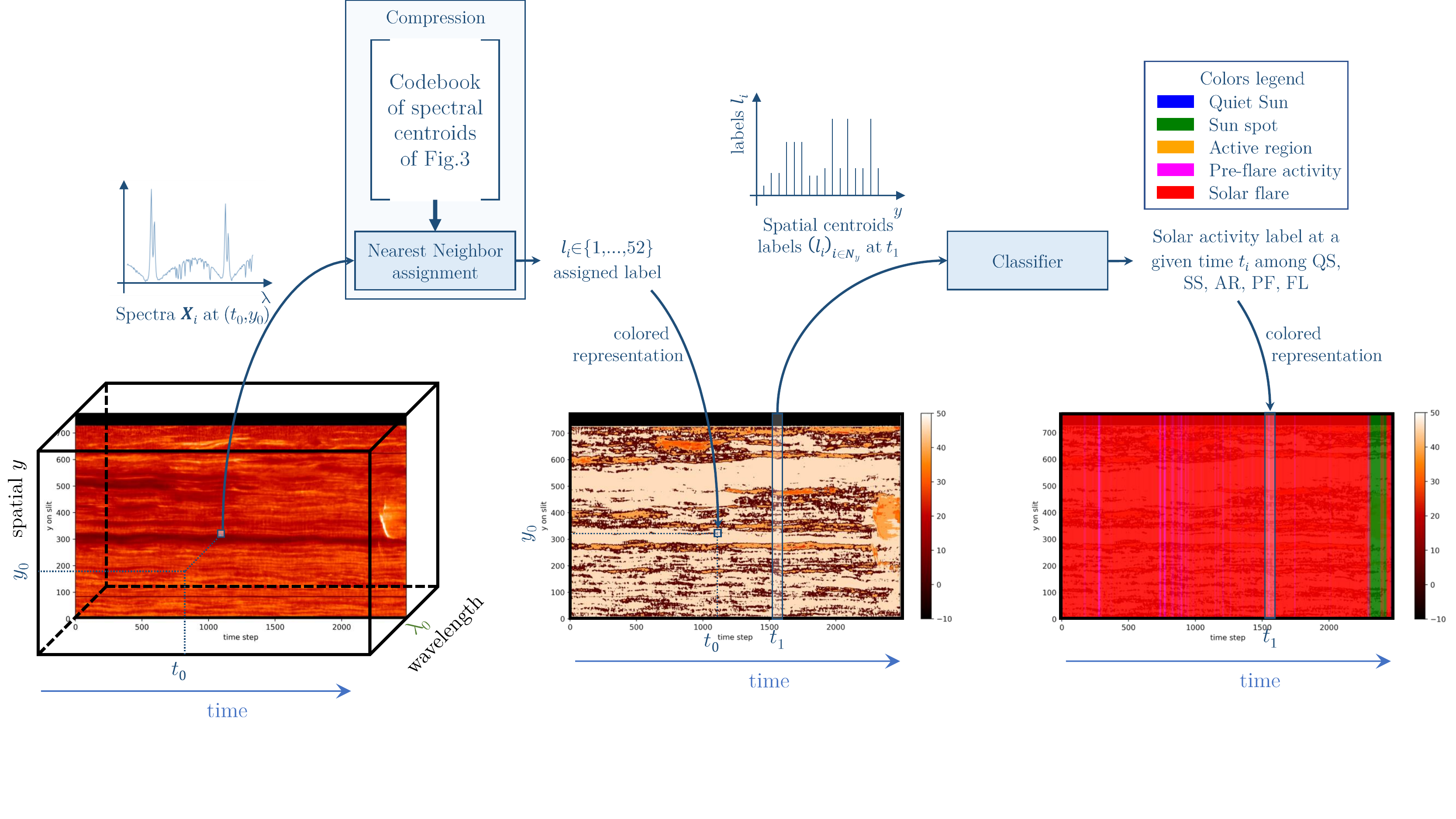}
\caption{Data compression and classification process. Left: Time evolution of spectra for an X-class flare observed on 10, September 2014\protect\footnotemark. Center: Time evolution of spectral clusters, with the 53 labels indicated by the corresponding color map. Black colored pixels are used to indicate missing or corrupt data. Right: Activity classification for this observation. Colors correspond to those on Fig.4}
\label{compress-class-process}
\end{figure*}

The PF dataset contains Mg II spectra from an active region, which eventually produces a solar flare, in contrast to the AR dataset which does not terminate in a solar flare. The PF spectra are collected from a 25 minute window before flare onset, as defined by the Geostationary Operational Environmental Satellite (GOES). Each observation has a different field of view, this is why we crop each photo only to the size of its 160 central pixels. We have only used the Mg II spectra if these observations. The spectra were clustered with the k-means algorithm described in \cite{Panos18}. The considered classification system based on the compressed data scheme is shown in Fig.\ref{compress-class-process}.

In the following section, we will investigate several types of classification algorithms and compare their performance. One such classifier is XGBoost proposed by \cite{Chen16}, which will occupy most of our attention, since it has recently been proposed as an alternative to deep classifiers, with relatively modest labeling requirements and more interpretable results.

\section{Gradient boosted trees method}
\label{Classifier_section}
The essence of classifier is to find a prediction of the value of observation according to a set of pre-defined prediction labels to minimize a loss function between the predicted label and assigned true label a.k.a. target label at the training set. In our case, the observation data that also serve as an input data to the classifier are represented as a sequence of numbers characterizing the centroids in the compressed IRIS observation frame. The classifier training based on the above minimization results in a set of hyperparamters of classifier. 

Decision trees are classification or regression schemes that are designed to iteratively and disjointedly group the space of the observations from the training dataset according to their attributes (predictors), ideally until each obtained group corresponds only to one decision (classification) class. Thus, when a new observation has to be classified, it is assigned to a single subgroup among those obtained in training, and labelled by the class corresponding to this subgroup.

The \emph{eXtreme Gradient Boost} or simply \emph{XGBoost} ML method is a gradient-boosted decision tree method introduced in \cite{Chen16}. Almost immediately it became very popular among data scientists, as it has revealed itself  both as a good classifier and regressor in different competitions. In astrophysics, it is also often used, for features restoring in large scale structure studies \cite{Calderon19,Tsizh20}, or for object selection and classification in observational catalogues by \cite{Jin19, Wang19}. For a short qualitative description one can consult \cite{Tsizh20}, while \cite{Jin19} contains a short quantitative description. We will summarize the goals and theory of the XGBoost method here.

XGBoost belongs to the group of decision trees ML techniques. Decision trees methods work in the following way. They generate a set of "weak" predictors based on individual trees each of which splits the training set in its own way using differences in features values. The name "weak" predictor comes from the fact that individually each such a predictor gives a very imprecise prediction. Every split is made in such a way that it forms the most homogeneous branches in terms of the target variables. The set of trees is diversified, as the initial split is made with different feature variables. Each tree ends up with a number of "leaves" and subcategories with almost the same types of instances. Having a set of trees to classify a new observation, "voting" is performed, during which each tree gives its own class prediction, with the final results depending on which class takes most "votes".

This algorithm is improved iteratively in the XGBoost method. The total objective function $\mathcal{L}$ of the whole tree set is expanded with an individual loss function of one of the weak predictors. The summation over the training set observations is performed with some weight coefficients multiplied with this single loss function. To find the latter ones, that contribute to the minimization of the total loss function the best, the gradient descent optimization is used. The gradient is calculated in a space of statistics on the total loss function, i.e., it entirely depends on the given population, and the difference of loss functions between two consecutive iterations is taken.

Mathematically, it can be formulated as follows. The objective function of a decision tree method is:
$$\mathcal{L} = \sum_{i=1}^M \ell(m_i, \hat{m}_i) + \sum_{k=1}^K \Omega (\mathfrak{F}_k).$$
Here $\hat{m}_i = \sum_{k=1}^K \mathfrak{F}_k(\bm{Z}_i)$ and $\mathfrak{F}_k(\bm{Z}_i)$ is a "weak predictor" for the separate tree of index $k$, $m_i$ is a true label applied to the compressed data $\bm{Z}_i$. $\ell(.,.)$ is the individual loss function and $\sum_{k=1}^K \Omega (\mathfrak{F}_k)$ is a regularization term. 
The summation $\sum_k$ is over the ensemble of the $K$ decision trees. The modification in evaluating the loss function iteratively, on the $t$-th step is:
$$\mathcal{L}^{(t)} = \sum_{i=1}^N [\ell(m_i, \hat{m}_i) + g_i \mathfrak{F}_t(\bm{Z}_i) + \frac{h_i}{2}\mathfrak{F}_t^2(\bm{Z}_i)] + \sum_{k=1}^K\Omega (\mathfrak{F}_k).$$
Here two terms of the first and second order of $f_t$ were added. The coefficients $g_i$ and $h_i$, as was mentioned before, are determined through gradient descent algorithm at each step:
$$g_i =  \partial_{\hat{m}^{(t-1)}} \ell(m_i, \hat{m}_i^{(t-1)}), \quad  h_i = \partial^2_{\hat{m}^{(t-1)}} \ell(m_i, \hat{m}_i^{(t-1)}).$$
The coefficients $g_i$ and $h_i$ are, in a sense, coefficients of Taylor expansion.

Usually after several tens or hundreds of iterations the algorithm converges to a solution providing some significant improvement over a simple decision tree method. 

\subsection{Application of XGBoost to solar spectra}
In our work, the feature variables of the ML algorithm are arrays of cluster labels $\bm{Z}$. The range of each cluster variable is $\left\{1,\dots,52\right\}$, meaning that there are 52 types of spectral shapes, and some of them, naturally are typical for solar flares, as pointed out in \cite{Panos18}. The value -10 was prescribed to those spectra that have been labeled either missing or corrupted by the pre-processing pipeline. Figure \ref{compress-class-process} shows a large sit-and-stare IRIS observation of an X-class flare on 10, September 2014. The evolution of the Mg II k spectra is pictured in Fig.\ref{compress-class-process}-Left. Note that this data is actually a 1D-image, as the spectrometer takes spectra over the entire slit.  The same observation in terms of the labels of each spectra is given in Fig.\ref{compress-class-process}-Center.\footnotetext{reference of the observation: $20140910\_112825\_3860259453$}
There is an abrupt increase of flux towards the end of the observation, which is related to a change of solar activity.  

\section{Results}
\label{Results_section}
\begin{figure}[thb]
\begin{subfigure}{.47\linewidth}
\centering\includegraphics[width=\textwidth, trim=60mm 0mm 50mm 0mm, clip]{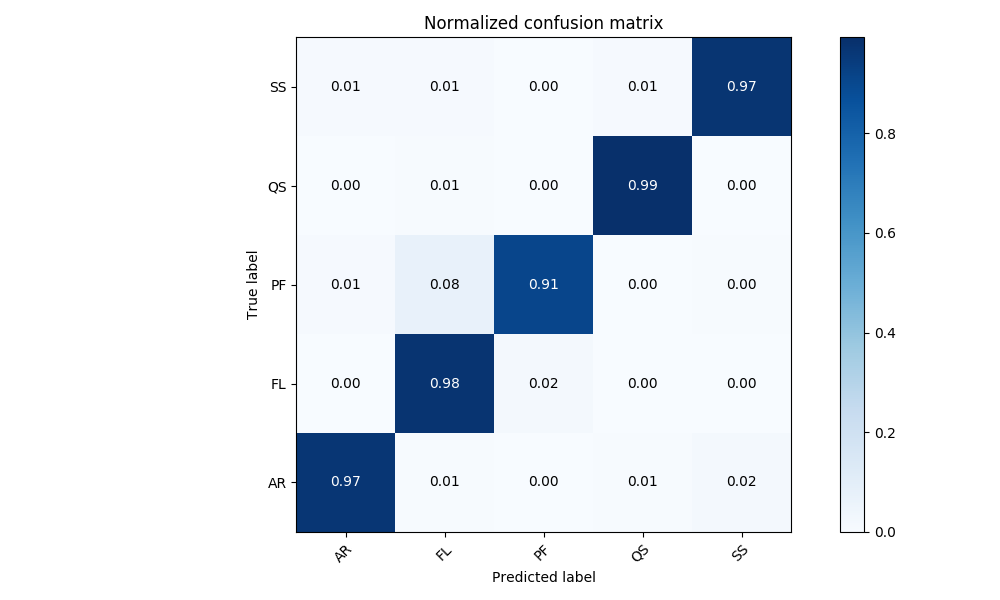}
\caption{On compressed data}
\label{subfig:cm-compressed}
\end{subfigure}
\begin{subfigure}{.47\linewidth}
\centering\includegraphics[width=\textwidth, trim=60mm 0mm 50mm 0mm, clip]{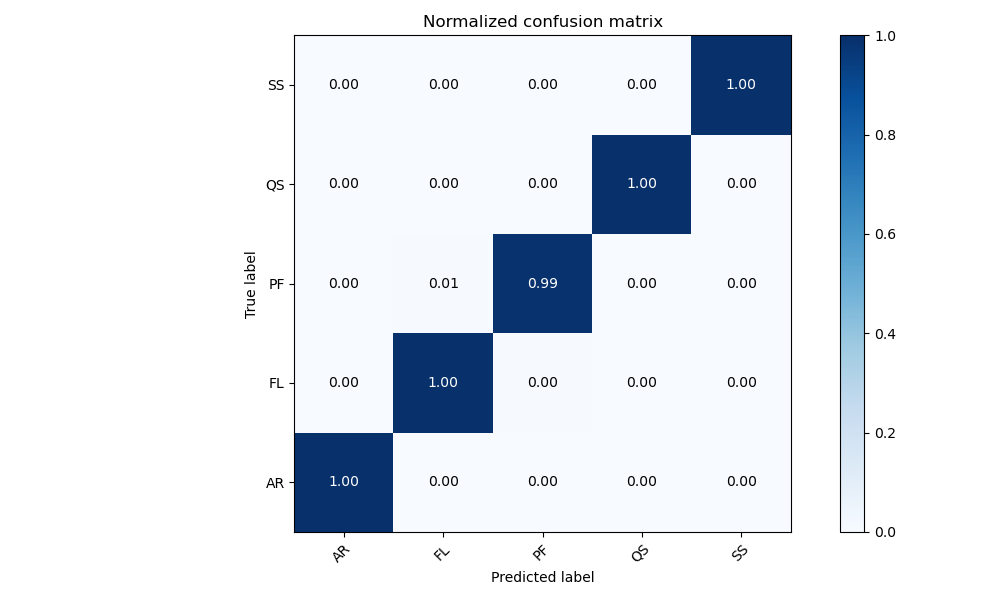}
\caption{On raw data}
\label{subfig:cm-raw}
\end{subfigure}
\caption{The resulting confusion matrices of activity classification using the XGBoost ML method. The confusion matrix counts, for each actual class, the percentages of predicted classes, and gives a nice visualisation of the misclassified cases. A perfect predictor has ones on the diagonal and zeros elsewhere. 0.00 and 1.00 displayed here are rounded up to 0.01 (1\%).}
\label{confmatr}
\end{figure}

After a grid search over the parameter space, we found the best achievable classification rate to be 95.77 $\%$, with a maximal tree depth of 18 (representing the constraint on number of "leaves") and number of estimators K set to 147, which corresponds to the number of weak predictors. The cross validation was performed with the $k$-folding method, with $k=5$. The confusion matrix is given in Fig. \ref{subfig:cm-compressed}.

We note that the strongest confusion exists between flare and pre-flare types of activity. This result is expected, as these types of solar activities share many physical commonalities. Quiet Sun, on the contrary, is almost never misclassified. One can find the code used for these computations at this\footnote{\href{https://github.com/DenisUllmann/Solar-activity-classification-based-on-Mg-II-spectra-towards-classification-on-compressed-data}{https://github.com/DenisUllmann/Solar-activity-classification-based-on-Mg-II-spectra-towards-classification-on-compressed-data} } public repository.

\subsection{Comparison to other methods}
We have performed the comparison of XGBoost with several classifiers of other families trained on the same compressed data:
\begin{itemize}
\item Convolutional neural network (CNN) with one convolutional, one max-pooling and one dense layer, trained for 300 epochs.
\item k-nearest neighbors (k-NN) with $k=6$.
\item Naive Bayes classifier.
\item Support vector machine (SVM) with a polynomial kernel.
\end{itemize}
The results are given in Table \ref{table2}. As was mentioned before, the best score was achieved with XGBoost method. The latter was performed with the \texttt{xgboost} package.
\begin{table}
\begin{center}
\begin{tabular}{cc} 
  \toprule
  \multicolumn{2}{c}{\bf On compressed data}\\
  \midrule
  ML classification method & Prediction score \\
  \midrule
  CNN & 0.883 \\
  k-NN & 0.871 \\
  Naive Bayes & 0.572 \\
  SVM &  0.638 \\
  XGBoost & {\bf 0.958} \\ 
  \bottomrule
  \toprule
  \multicolumn{2}{c}{\bf On raw data}\\
  \midrule
  ML classification method & Prediction score \\
  \midrule
  XGBoost & {\bf 0.976} \\ 
  \bottomrule
\end{tabular}
\end{center}
\caption{Results for different ML methods on compressed data and comparison with the prediction on raw data.}
\label{table2}
\end{table}

It is interesting to point out that the XGBoost method has achieved better results than convolutional networks, which are popular today for image classification. We believe the main reason for this is that we deal with one-dimensional signals that have constrained number of possible configurations. XGBoost has also out-performed methods based on distances in the feature space, such as k-NN and SVM, since the number of features is sufficiently large.

\subsection{Classification with uncompressed data}
In order to quantify the drop in classification accuracy induced by our proposed classification on compressed data, we applied the XGBoost method to raw spectral data, without having the data reduction described in the middle part of Figure \ref{compress-class-process}. After a grid search over the parameter space, we found the best achievable classification rate to be 97.63$\%$, in comparison to the 95.77 $\%$ achieved on the compressed data. This result was obtained with a maximal tree depth of 9 and number of estimators $K = 185$. The cross validation was performed once again with the $k$-folding method, retaining the same number of folds as before. The confusion matrix is given in Fig. \ref{subfig:cm-raw}. The confusion matrix is compatible to that of the compressed data, but with higher aggregate scores. 

\subsection{Independent tests and limitations}
We tested the classification scheme on a preflare observation shown in Figure \ref{compress-class-process}-Right. The ML model weighs the compressed groups for each slice in time and assigns a corresponding class label. The classification undergoes a transformation from red, indicating flare, to green indicating Sunspot, precisely at the point where a substantial increase in flux occurs (as indicated in the left hand panel). Although the observation is taken before the onset of a large X1.6-class flare, there is a small C-class flare that occurs within the same region several minutes earlier, which may explain the categorization of the majority of the observation as flaring. The Sunspot categorization might be explained by the production of large numbers of single-peaked spectra falling within spectral group 46 in Figure \ref{Centroids}. This follows from the Mg II lines propensity to re-couple to the Plank function over intense, hot solar regions, allowing the source function to continue increasing with height.

There are several limitations that must be addressed here. These limitations do not undermine our methods, but rather the extent to which the models would be applicable for solar activity classifications at different radius's away from disk center. 

The training data set was selected from the central part of the solar disk. One can expect, that in order to perform accurately on the limb, new centroids specific to this region should be generated as priors for the ML model. With the current disk-center centroids, the model would be less efficient in its class predictions at the edges of the disk, since center to limb variations produce different spectral shapes (limb darkening), and the dictionary used for compression would therefore be highly incompatible with solar limb observations. Finally, the dictionary in Figure \ref{Centroids} was prioritized for flare observations, an extended dictionary that gives equal wight to each of the five solar regions would lead to stronger and more interpretable results. 

\section{Conclusion}
\label{Conclusion_section}
We have successfully demonstrated how a carefully guided lossy compression scheme (k-means), coupled with a scalable classifier (XGBoost), can achieve high predictions scores of $95\%$ , when applied to the task of organising five different categories of solar activity based on Mg II spectra alone. Furthermore, the results of the classifier applied to the compressed data are comparable to those applied to the raw uncompressed data. This implies that the compression scheme captures most of the relevant features for the task at hand, and that the large gains in compression come at a small price to the accuracy. This particular architecture can be easily  scaled to much larger data sets with a relatively low time complexity. It takes only 30-60 minutes for the dateset we used in this work to be processed and ML model training and only few seconds to provide the classification of an observation with a thousand frames on a typical office desktop computer.

We show that the entire raw data captured by a spacecraft is not necessary to perform interesting data analysis, as those considered in the paper, and we demonstrate that very high rates of lossy compression still provide essentially useful information for accurate data analysis and interpretation. Moreover, the Information Bottleneck (IB) principle suggests an autoencoder scheme to achieve a classification or a regression, such that the raw data has to be encoded into a latent representation which is a compressed version of the raw data, and then decoded into the targeted information. Once the encoder and maybe the decoder as well are known, it is enough to store on-board just the compressed data given by the encoder, or even the targeted information given by the decoder while the last one saves even more space. In our experiments the output of the decoder is represented by the the solar activity encoded into 5 labels that requires just  3 bits. It should be noted, that when several types of information are required from the raw data, different pairs of encoders - decoders may have to be learned and then different compressed representations of the same raw data may have to be stored on-board before a transmission to the earth. Overall, we demonstrate a way how to reduce and optimize the amount of information stored on-board and thus to increase the number of analysis, functionalities and missions that can be held by the spacecraft.

\section{Acknowledgements}
This work was funded by Abto Software LLC. Denis Ullmann and Brandon Panos were supported by the Swiss National Science Foundation (SNSF) in the scope of the NRP75 project number 407540\textunderscore167158. We would like to thank Dr. O. Makoveychuk and Dr. M. Stodilka for fruitful discussions and valuable advises during our work. We are also thankful to Prof. L. Kleint, Prof. M. Melchior and Dr. C. Huwyler for their help and discussions about the interpretability of data provided by IRIS.

\end{document}